\documentclass[aip,cha,reprint]{revtex4-1}

\usepackage{graphicx}
\usepackage{SIunits}
\usepackage{hyperref}
\usepackage{amsmath}

\newcommand{\geff}{g_{\mathrm{eff}}}

\begin{document}

\title{High cooperativity coupling between a phosphorus donor spin ensemble and a superconducting microwave resonator}

\author{Christoph W. Zollitsch}
\email{christoph.zollitsch@wmi.badw.de}
\affiliation{Walther-Mei\ss{}ner-Institut, Bayerische Akademie der Wissenschaften, 85748 Garching, Germany}
\affiliation{Physik-Department, Technische Universit\"at M\"unchen, 85748 Garching, Germany}

\author{Kai Mueller}
\affiliation{Walther-Mei\ss{}ner-Institut, Bayerische Akademie der Wissenschaften, 85748 Garching, Germany}
\affiliation{Physik-Department, Technische Universit\"at M\"unchen, 85748 Garching, Germany}

\author{{David P. Franke}}
\affiliation{Physik-Department, Technische Universit\"at M\"unchen, 85748 Garching, Germany}
\affiliation{Walter Schottky Institut, Technische Universit\"at M\"unchen, 85748 Garching, Germany}

\author{{Sebastian T. B. Goennenwein}}
\affiliation{Walther-Mei\ss{}ner-Institut, Bayerische Akademie der Wissenschaften, 85748 Garching, Germany}
\affiliation{Nanosystems Initiative Munich, 80799 M\"{u}nchen, Germany}

\author{Martin S. Brandt}
\affiliation{Physik-Department, Technische Universit\"at M\"unchen, 85748 Garching, Germany}
\affiliation{Walter Schottky Institut, Technische Universit\"at M\"unchen, 85748 Garching, Germany}

\author{Rudolf Gross}
\affiliation{Walther-Mei\ss{}ner-Institut, Bayerische Akademie der Wissenschaften, 85748 Garching, Germany}
\affiliation{Physik-Department, Technische Universit\"at M\"unchen, 85748 Garching, Germany}
\affiliation{Nanosystems Initiative Munich, 80799 M\"{u}nchen, Germany}

\author{Hans Huebl}
\email{huebl@wmi.badw.de}
\affiliation{Walther-Mei\ss{}ner-Institut, Bayerische Akademie der Wissenschaften, 85748 Garching, Germany}
\affiliation{Nanosystems Initiative Munich, 80799 M\"{u}nchen, Germany}

\date{\today}

\begin{abstract}

We investigate the coupling of an ensemble of phosphorus donors in an isotopically purified $^{28}$Si host lattice interacting with a superconducting coplanar waveguide resonator. The microwave transmission spectrum of the resonator shows a normal mode splitting characteristic for high cooperativity. The evaluated collective coupling strength $\geff$ is of the same order as the loss rate of the spin system $\gamma$, indicating the onset of strong coupling. We develop a statistical model to describe the influence of temperature on the coupling strength from $50\,\milli\kelvin$ to $3.5\,\kelvin$ and find a scaling of the coupling strength with the square root of the number of thermally polarized spins.

\end{abstract}

\pacs{42.50.Pq, 76.30.Mi, 85.25.-j, 85.70.-w}

\keywords{silicon, phosphorus donors, strong coupling, high cooperativity, quantum memory}

\maketitle
Quantum memories complement fast quantum processing elements in solid-state based quantum information processing. Typically these two elements come with opposing requirements as long storage times depend on systems well isolated from their environment \cite{Tyryshkin2012, Saeedi2013} while fast operating elements rely heavily on strong interactions amongst themselves as well to control circuits \cite{Schoelkopf2008, Hofheinz2009}. One approach to reconcile these contradictory requirements is to strongly couple both memory and operating elements in a hybrid circuit e.g. via a quantum bus system \cite{Xiang2013}. Promising candidates for storage elements are spin ensembles due to their long coherence times, while for quantum processing elements superconducting quantum circuits are well established. Notably, both elements operate in the same GHz frequency regime.

A multitude of experiments demonstrated the preparation of quantum states using superconducting circuits \cite{Schoelkopf2008, Xiang2013, Hofheinz2009, Niemczyk2010, Wallraff2004}. Complementary, the implementation of (classical) information storage and retrieval protocols were realized for natural spin systems such as nitrogen vacancy (NV) centers in diamond \cite{Kubo2012, Grezes2014} and phosphorus donors in silicon \cite{Wu2010}. Additionally, strong coupling between microwave resonators and spin ensembles was demonstrated for NV centers \cite{Kubo2010, Kubo2011, Sandner2012, Putz2014} and P1 centers \cite{Ranjan2013} in diamond, Cr impurities in sapphire \cite{Schuster2010}, and erbium centers in Y$_2$SiO$_5$ \cite{Bushev2011} as well as for exchange-coupled magnetic systems \cite{Huebl2013, Tabuchi2014, Zhang2014, Cao2015, Bai2015}. However, a demonstration of strong coupling between phosphorus donors in silicon and a microwave resonator is still lacking, although phosphorus dopants in isotopically 
purified $^{28}$Si are a benchmark system due to their exceptionally long electron and nuclear coherence times exceeding $0.5\,\second$  and $39\,\minute$, respectively \cite{Muhonen2014, Tyryshkin2012, Saeedi2013}. Furthermore, to inhibit thermal noise to interfere with the quantum information the circuit must be operated at millikelvin temperatures, a temperature regime rather unexplored for phosphorus donors in silicon \cite{Sigillito2014}.

In this letter, we investigate the interaction of a superconducting niobium coplanar waveguide resonator (CPWR) with an ensemble of phosphorus donors in an isotopically enriched $^{28}$Si host in the millikelvin temperature regime. We determine the coupling rate $\geff$ between the microwave resonator and the electron spin ensemble and find that it is of similar size as the spin ensembles dephasing rate $\gamma$ at the lowest temperatures used, indicating a regime of high cooperativity. We further investigate the coupling rate between the CPWR and the phosphorus spins as a function of temperature up to $3.5\,\kelvin$ and can quantitatively describe the data in terms of the thermal spin polarization $P(T)$.

For a coupled system consisting of a spin ensemble and a microwave resonator, three (partly overlapping) coupling regimes can be distinguished: (i) The weak coupling regime: Here, the collective coupling rate $\geff$ between the microwave resonator and the spin ensemble is smaller than their individual loss rates $\kappa_0$ and $\gamma$, respectively. (ii) The regime of high cooperativity with the cooperativity $C=\geff^2/(\kappa_0 \gamma)>1$: Here, the interaction rate $\geff$ causes a level repulsion of the eigenstates of the spin ensemble and the microwave cavity to an extend that a double peak becomes spectroscopically visible. (iii) The strong coupling regime: Here, the coupling rate $\geff$ exceeds both individual loss rates $\kappa_0$ and $\gamma$ allowing for a coherent exchange of (quantum) excitations between the subsystems \cite{Walls2008}. For a spin ensemble coupled to a microwave resonator the collective coupling rate $\geff=g_0 \sqrt{N P(T)}$, where $N$ is the number of spins in the cavity, $P(T)$ is the thermal spin polarization factor, and $g_0=(g_{\mathrm{s}}\mu_{\mathrm{B}}/2\hbar)\sqrt{\mu_0\hbar\omega_{\mathrm{r}}/2V_{\mathrm{m}}}$ is the coupling rate of a single spin to the vacuum field of the microwave resonator \cite{Sandner2012,Huebl2013,Wesenberg2009}. Here, $g_{\mathrm{s}}$ is the electron g-factor, $\mu_{\mathrm{B}}$ the electron Bohr magneton and $\omega_{\mathrm{r}}$ and $V_{\mathrm{m}}$ are the resonance frequency and the mode volume of the resonator, respectively.

\begin{figure}[!ht]
 \includegraphics[]{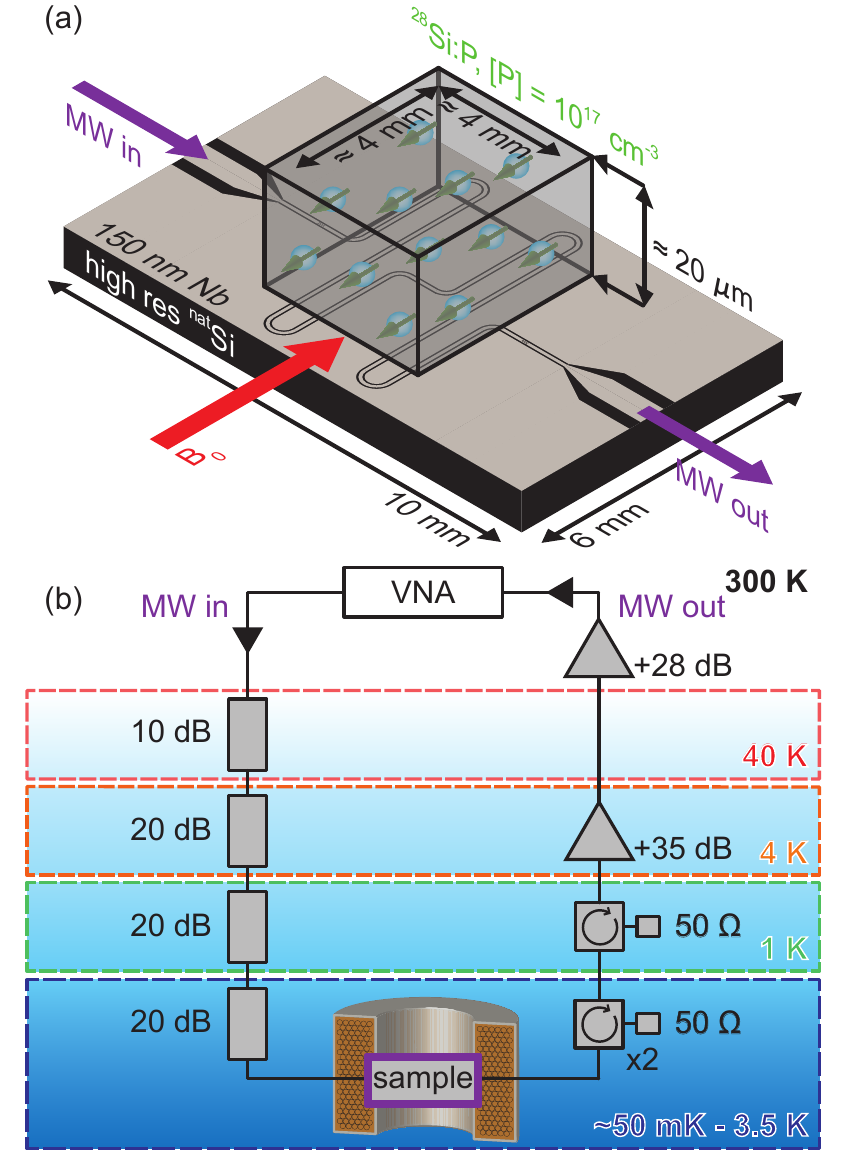}
 \caption{(a) Schematic of the superconducting CPWR with a phosphorus-doped $^{28}$Si crystal positioned on top of the resonator structure. The $^{28}$Si crystal is held in place by a piece of high resistivity silicon wafer (not shown). The static magnetic field $B_0$ is oriented in parallel to the superconducting thin film. (b) Schematic of the dry dilution refrigerator and the microwave spectroscopy setup. To inhibit thermal photons exciting the sample the microwave input and output line is equipped with attenuators and circulators, respectively. Additionally, the CPWR output is amplified by a cryogenic HEMT amplifier and a room temperature low noise pre-amplifier. The sample is placed inside a superconducting solenoid. \label{fig:setup}}
\end{figure}

To study the interaction between a phosphorus spin ensemble and a superconducting microwave resonator, we place a $20\,\micro\meter$ thick isotopically enriched $^{28}$Si crystal doped with phosphorus atoms onto a CPWR. The $^{28}$Si crystal is arranged on the CPWR such that it covers most of the resonator structure, as schematically shown in Fig.\,\ref{fig:setup}\,(a).  The $^{28}$Si host material features a residual $^{29}$Si nuclei concentration of 0.1$\,\%$ and is homogeneously doped with $[\mathrm{P}]=10^{17}\,\centi\meter^{-3}$. For details regarding the CPWR and the mounting of the $^{28}$Si crystal please refer to the supplemental material \cite{supplement}. 
The first three modes of the superconducting coplanar $\lambda$/2 transmission line resonator have eigenfrequencies of $\omega_0/2\pi=2.47\,\giga\hertz$, $\omega_1/2\pi=4.94\,\giga\hertz$ and $\omega_2/2\pi=7.39\,\giga\hertz$ at $T=50\,\milli\kelvin$ and a static magnetic field of $B_0=0\,\tesla$. The corresponding quality factors are $Q_0=21200$, $Q_1=7172$ and $Q_2=3861$ for an input power at the resonator of $-134\,\mathrm{dBm}$. Note that the observed decrease of the quality factor with mode index is characteristic for this type of resonator design \cite{Goeppl2008}.
To study the resonance frequency evolution of the microwave resonator and the spin ensemble, we perform microwave transmission spectroscopy as a function of the externally applied magnetic field $B_0$, employing a commercial vector network analyzer (VNA). For this we aligned the orientation of $B_0$ parallel to the CPWR surface to maintain the high quality factor \cite{Wallace1991}. The experiments are performed as a function of temperature between $50\,\milli\kelvin$ and $3.5\,\kelvin$ using a dry dilution refrigerator. We attenuate the input signal line and use circulators for the output lines to isolate the device from thermal photons as schematically depicted in Fig.\,\ref{fig:setup}\,(b).

\begin{figure}[b]
 \includegraphics[]{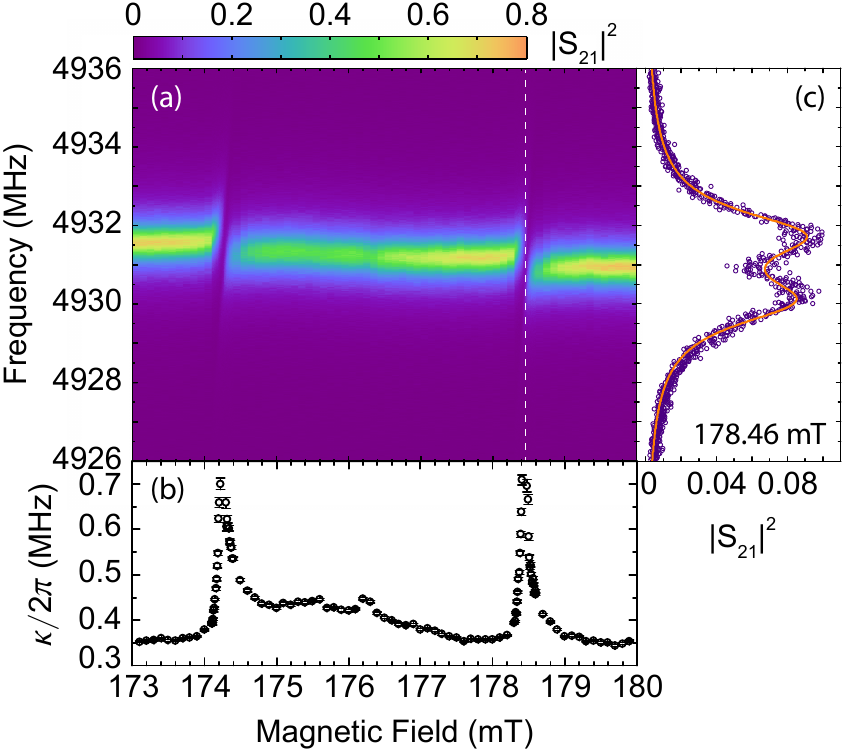}
 \caption{Spectroscopy data of the superconducting microwave resonator coupled to the phosphorus spin ensemble taken at a temperature of $50\,\milli\kelvin$ and an input microwave power of $40\,\atto\watt$. a) Color encoded microwave transmission data as a function of the probe frequency and the static magnetic field $B_0$. The central line of high transmission indicates the resonance frequency of the CPWR. At $B_{0,1}=174.27\,\milli\tesla$ and $B_{0,2}=178.46\,\milli\tesla$ the resonator and the spin ensemble are in resonance resulting in a reduced transmission. These resonances are separated by $4.19\,\milli\tesla$, agreeing very well with the characteristic hyperfine splitting of phosphorus donors in silicon. b) Effective linewidth (circles with error bars) of the resonator $\kappa$ determined by a Lorentzian fit to the data in (a). Here, additional increases of $\kappa$ are attributed to dangling bond defects P$_b$ and P$_2$ dimers observed at $B_0=175.5\,\milli\tesla$ and $B_0=176.3\,\milli\tesla$, respectively. c) Resonator transmission $|\mathrm{S}_{21}|^2$ (circles) at $B_0=178.46\,\milli\tesla$ (white dashed line in (a)), the resonance field of the high field spin transition. The resonator mode exhibits the characteristic normal mode splitting resulting in two individual transmission peaks separated by $2\geff$. The solid orange line represents a fit using (\ref{eqn:inout}). \label{fig:spectrum}}
\end{figure}

Figure\,\ref{fig:spectrum}\,(a) shows the microwave transmission spectroscopy data of the coupled spin ensemble\,/\,microwave resonator system as a function of the magnetic field for a temperature of $T=50\,\milli\kelvin$. The spectrum is recorded using a microwave power of $-134\,\mathrm{dBm}$ or $40\,\atto\watt$ at the input of the resonator \cite{photon_comment}. The spectrum shows the characteristic high transmission of the microwave resonator at its first harmonic with $\omega_{\mathrm{r}}/2\pi = 4.931\,\giga\hertz$. Additionally, we observe a reduction of the microwave transmission $|\mathrm{S}_{21}|^2$ through the resonator by a factor of ten at the magnetic fields $B_{0,1}=174.27\,\milli\tesla$ and $B_{0,2}=178.46\,\milli\tesla$, where the precession frequency of the spin system matches the resonance frequency of the CPWR. The field separation of those two resonances is $4.19\,\
milli\tesla$ and is in very good agreement with the characteristic hyperfine splitting of phosphorus donors in silicon \cite{Feher1959}.

For a coarse analysis of the data, we fit a Lorentzian function to the $|\mathrm{S}_{21}|^2$ data for each magnetic field point. Hereby, we gain information about the effective half width at half maximum $\kappa$ of the microwave resonator shown in Fig.\,\ref{fig:spectrum}\,(b) which directly corresponds to the information obtained in a conventional electron spin resonance (ESR) experiment \cite{Poole1997}. Besides the two phosphorus resonances, we find a broad line at $B_0=175.5\,\milli\tesla$ which is compatible with the signature expected for dangling bond defects P$_{\mathrm{db}}$ at the Si/SiO$_2$ interface at the sample surface \cite{Stesmans1998}. Additionally, the central line at $B_0=176.3\,\milli\tesla$ is attributed to exchange coupled P$_2$ dimers \cite{Feher1955}. Their presence is expected for the high P doping concentration of our sample \cite{Cullis1975}.

To quantify the collective coupling $\geff$ of the (non-interacting) donor spins to the resonator photon field we investigated the resonator transmission in the field region of the two isolated phosphorus donor electron spin resonances. Figure\,\ref{fig:spectrum}\,(c) shows the microwave transmission $|\mathrm{S}_{21}|^2$ for the high field spin transition at $B_0=178.46\,\milli\tesla$. We find that the microwave response exhibits a normal mode splitting, indicative for the high cooperativity regime. The separation of the two transmission peaks corresponds to $2\geff$ and thus allows a direct determination of the collective coupling rate. For a more detailed analysis of the situation we fit an input-output model to our transmission data (orange line in Fig.\,\ref{fig:spectrum}\,(c)) \cite{Clerk2010, Schuster2010, Huebl2013}
\begin{equation}
 \left|\mathrm{S}_{21}\right|^2 = \bigg| \frac{\kappa_{\mathrm{c}}}{ i\left( \omega - \omega_{\mathrm{r}} \right) - \kappa_0 + \displaystyle\sum_{n=1}^{2} \frac{\geff^2}{ i\frac{g_{\mathrm{s}} \mu_{\mathrm{B}}}{\hbar}\left( B_0 - B_{0,n} \right) - \gamma } } \bigg|^2.
 \label{eqn:inout}
\end{equation}
\\
Here, $\kappa_{\mathrm{c}}$ represents the external coupling rate of the CPWR to the external microwave circuit. From the fit to the data outside the electron spin resonance we extract a resonator loss rate of $\kappa_0/2\pi=370\,\kilo\hertz$. Taken together with the resonator frequency of $\omega_{\mathrm{r}}/2\pi=4.931\,\giga\hertz$ this corresponds to a quality factor of $Q \approx 6600$, which is still high considering the static magnetic field of more than $170\,\milli\tesla$. Futhermore, the calibrated data allows to determine the external and the internal quality factor to $Q_{\mathrm{ext}}=9793$ and $Q_{\mathrm{int}}=20857$, respectively, indicating that the microwave resonator operates in the overcoupled regime \cite{Goeppl2008}. Analyzing the data from the resonance fields $B_{0,1}$ and $B_{0,2}$, we find for the loss rate of the spin ensemble $\gamma/2\pi=1.38 \text{ and } 1.40\,\mega\hertz$ and for the collective coupling rate  $g_{\mathrm{eff}}/2\pi=1.13 \text{ and } 1.07\,\mega\hertz$, respectively. Thus, the coupled system is in the high cooperativity regime for both resonances as $C=2.5 \text{ and } 2.0 \text{, which are both } >1$.

In the following, we compare the experimentally extracted collective coupling rate $\geff$ with theory. For a first estimate, we use $g_{\mathrm{eff,th}}=(g_\mathrm{s} \mu_\mathrm{B}/2\hbar) \sqrt{\mu_0 \rho \hbar \omega_{\mathrm{r}} \eta/2}$. (cf. Ref.\,\onlinecite{Huebl2013}) and obtain $g_{\mathrm{eff,th}}/2\pi=3.17\,\mega\hertz$. For this, we assume that (i) the upper half of the resonator mode volume is completely filled with the spin ensemble corresponding to a filling factor of $\eta=1/2$, that (ii) the spin density is $\rho=(1/2)\times 1 \times 10^{17}\,\centi\meter^{-3}$, because only half of the spins contribute to each of the two ESR transitions, and that (iii) only one circularly polarized component of the magnetic vacuum field interacts with the ESR transition \cite{coupling_comment}. This overestimates the experimental observation of $\geff$, which we attribute to a finite gap $d$ between the CPWR and the $^{28}$Si crystal.
We quantify this finite gap $d$ by calculating $\geff$ numerically. For this, we compute the spatially dependent individual single spin coupling rate $g_0$ taking the spatial microwave magnetic field distribution $B_1$ of the CPWR into account \cite{Simons1982}. We then obtain the effective coupling rate via $\geff=\sqrt{\sum_i |g_{0,i}|^2}$ (cf. Ref.\,\onlinecite{Sandner2012}). For a gap  of $d=12.5\,\micro\meter$, we obtain $\geff/2\pi=1.10\,\mega\hertz$, which is in good agreement with the experimentally determined values (for details see Ref.\,\onlinecite{supplement}). The corresponding gap is attributed to the absence of special CPWR and $^{28}$Si crystal cleaning procedures prior to mounting the spin ensemble.

The extracted spin ensemble loss rate $\gamma$ translates to a magnetic full width at half maximum (FWHM) linewidth of  $\Delta B=98.52 \text{ and } 99.80\,\micro\tesla$ for the low and high field resonance, respectively. This is a factor of 10 broader than expected for phosphorus donors in $^{28}$Si with the residual $^{29}$Si of the sample used \cite{Tyryshkin2012, Morishita2011, Gordon1958}. We attribute this discrepancy to inhomogeneities in the static magnetic field $B_0$ provided by the superconducting magnet. Reference experiments with the same crystal in a commercial Bruker electron spin resonance spectrometer performed at a temperature of $8\,\kelvin$ yield a linewidth of $\Delta B_{\mathrm{Bruker}} = 37.7\,\micro\tesla$. The latter corresponds to a $\gamma_{\mathrm{Bruker}}/2\pi =530\,\kilo\hertz$ indicating that $\gamma$ is increased by inhomogeneous effects, which limits the cooperativity of the hybrid system \cite{supplement}.
Therefore, considering $\gamma_{\mathrm{Bruker}}$ instead of $\gamma$ places the coupled system in the strong coupling regime and correspondingly results in a cooperativity of $C=6.5 \text{ and } 5.8$ for both electron spin resonances.

\begin{figure}[t]
 \includegraphics[]{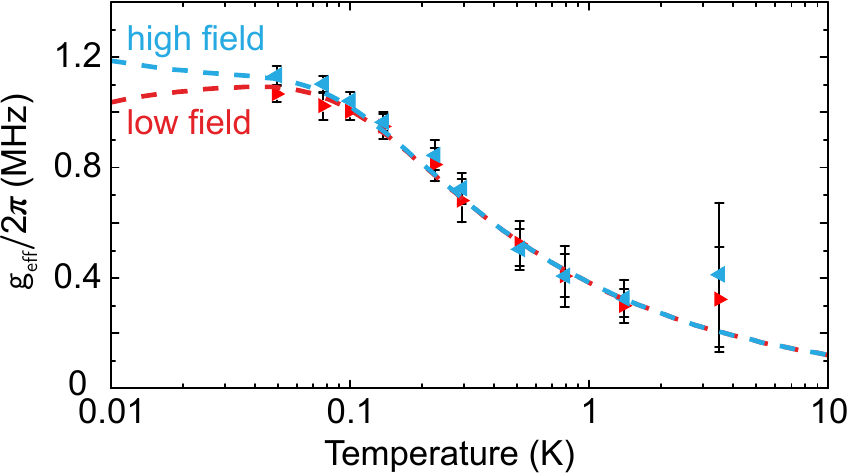}
 \caption{Collective coupling $\geff$ as a function of logarithmic scaled temperature for both the low field (blue triangles) and high field (red triangles) spin transition. For low temperatures there is a finite difference in the coupling rates, which is attributed to the thermal nuclear spin polarization. For high temperatures ($T>100\,\milli\kelvin$) we find the characteristic $T^{-1/2}$ behaviour expected for a $S=1/2$ spin system. The dependence is well described by our model (dashed lines), taking all eigenenergies of the phosphorus donor ($S=1/2$, $I=1/2$) into account. \label{fig:couplings}}
\end{figure}

Next, we investigate the collective coupling strength $\geff$ as a function of temperature up to $3.5\,\kelvin$ for both the low field and high field resonance as shown in Fig.\,\ref{fig:couplings}. With increasing temperature the coupling strength reduces from its maximum at $50\,\milli\kelvin$. For temperatures above $100\,\milli\kelvin$ the coupling to both resonances follows the same temperature dependence. Below $100\,\milli\kelvin$ we observe a finite difference in the coupling strength of the two resonances with the low field resonance having the higher $\geff$. This behavior is understood as $\geff = g_0 \sqrt{N P(T)}$ scales with the thermal polarization $P(T)$ of the respective spin resonance transition \cite{Sandner2012}. $P(T)$ can be derived from the thermal population of the individual energy levels of the phosphorus donor spin system with electron spin $S=1/2$ and nuclear spin $I=1/2$. \cite{supplement} Our data corroborates the expected theoretical behavior as indicated by the dashed lines in Fig.\,\ref{fig:couplings}. Below $100\,\milli\kelvin$, the electron spin system is fully polarized and the nuclear thermal polarization gives rise to a difference in the collective coupling constants for the two spin resonance transitions. In contrast, above $100\,\milli\kelvin$, the thermal nuclear polarization is negligible and the characteristic $T^{-1/2}$ dependence expected for the thermal electron polarization of a $S=1/2$ system dominates the dependence of the coupling constants.

In summary, we demonstrated a phosphorus donor spin ensemble coupled to a superconducting coplanar microwave resonator in the high cooperativity regime. The coupling is well described by an input-output formalism allowing to extract a collective coupling strength of $1.13\,\mega\hertz$ and $1.07\,\mega\hertz$ for the two phosphorus resonance transitions, which are corroborated quantitatively by a theoretical estimate. The spin dephasing rate $\gamma$ determined is limited experimentally by the magnetic field inhomogeneities of the superconducting magnet. In addition, we provide a statistical model to predict the temperature dependence of the collective coupling strength by taking the thermal spin polarization of both the electron and nuclear spins of the phosphorus donors into account and find excellent agreement with the experimental data.

This work is financially supported by the Deutsche Forschungsgemeinschaft through SFB 631 C3, A3 and the Priority Programm SPP 1601 (HU 1896/2-1).

\pagebreak
\widetext
\begin{center}
\textbf{\large High cooperativity coupling between a phosphorus donor spin ensemble and a superconducting microwave resonator: Supplemental Information}
\end{center}

\setcounter{equation}{0}
\setcounter{figure}{0}
\setcounter{table}{0}
\setcounter{page}{1}
\makeatletter
\renewcommand{\theequation}{S\arabic{equation}}
\renewcommand{\thefigure}{S\arabic{figure}}
\renewcommand{\bibnumfmt}[1]{[S#1]}
\renewcommand{\citenumfont}[1]{S#1}

\section{Superconducting Coplanar Waveguide Resonator}

\begin{figure}[!ht]
 \includegraphics[]{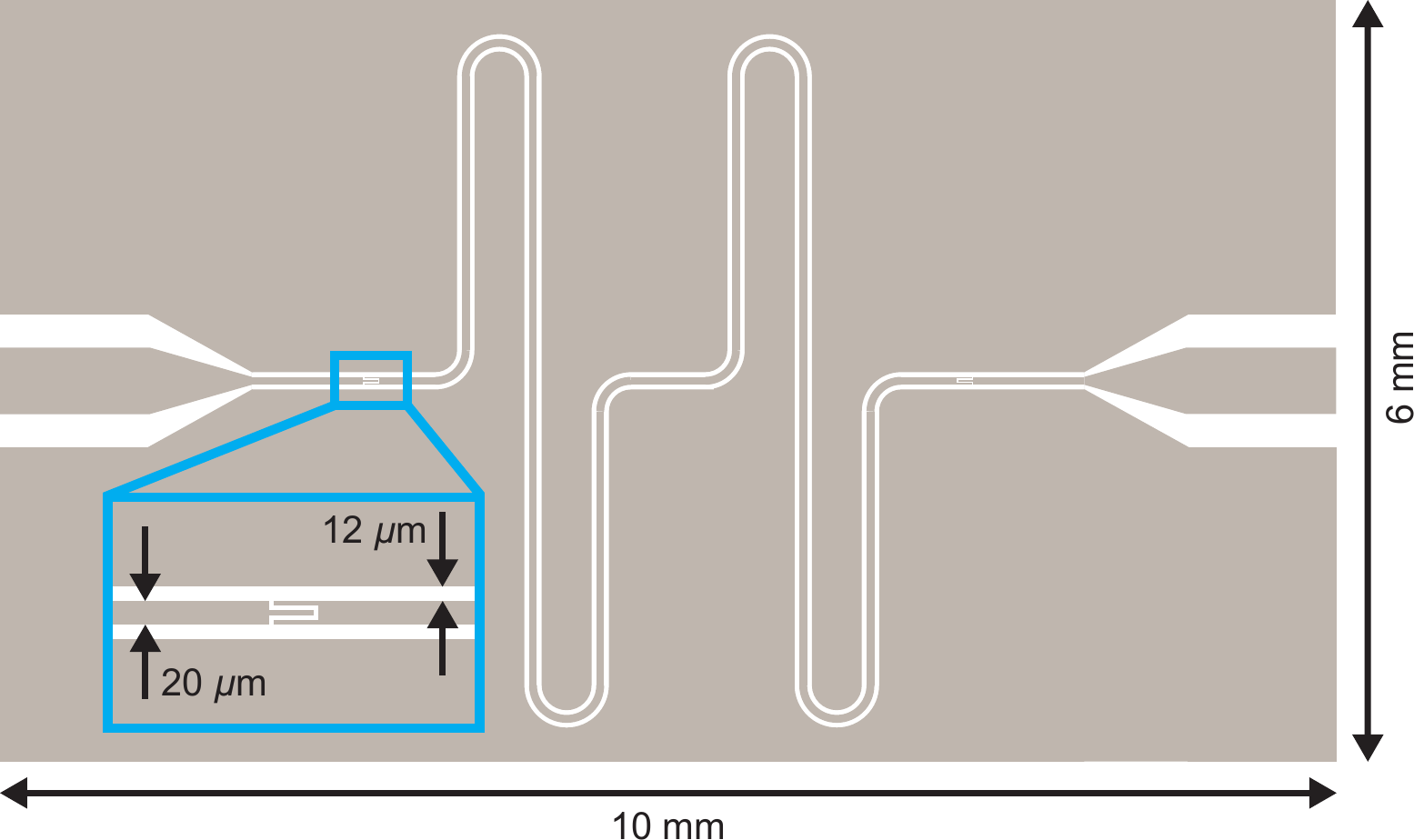}
 \caption{Schematic layout of the coplanar waveguide resonator. The inset shows a magnification of one of the two coupling capacitors of the resonator. The center conductor width is $20\,\micro\meter$ and the gap width is $12\,\micro\meter$. The resonator structure has a total length of $23\,\milli\meter$. \label{S1}}
 \end{figure}

For the fabrication of the superconducting coplanar waveguide resonators (CPWR), we start with a $6\times 10\,\milli\meter^2$ high resistivity ($> 3\,\kilo\ohm\centi\meter$) Si substrate with a natural isotope composition on which we deposit a $150\,\nano\meter$ thick niobium layer using magnetron sputtering. We define the CPWR structure using optical lithography and reactive ion etching. Finally, we obtain a transmission coplanar waveguide resonator structure as depicted in Fig.\,\ref{S1}. The impedance of the coplanar waveguide resonator is designed to $Z=50\,\ohm$, which is obtained for a width of the center conductor of $20\,\micro\meter$ and a gap between the center conductor and the ground plane of $12\,\micro\meter$.

\section{Sample Preparation and Mounting }

For the spin ensemble we use phosphorus donors with a donor concentration of $[\mathrm{P}]=10^{17}\,\centi\meter^{-3}$ contained in a $20\,\micro\meter$ thin isotopically enriched $^{28}$Si crystal. The here used isotopically purified $^{28}$Si crystal was initially grown by chemical vapor deposition on top of a silicon substrate with natural Si isotope composition ($^{\mathrm{nat}}$Si). This substrate is boron-doped to a level above the metal insulator transition and thus has a finite conductance at low temperatures. The conducting substrate would induce microwave losses and therefore, we have removed the substrate by mechanical polishing. The resulting thin $^{28}$Si crystals are placed onto the CPWR and held in position by an additional piece of high resistivity $^{\mathrm{nat}}$Si substrate. The latter is cemented onto the resonator. Note that the thin $^{28}$Si crystals extend over almost the entire area of the coplanar microwave resonator as sketched in Fig.\,1\,(a) of the main text. This spacially 
extended overlap allows for coupling the spin system to the fundamental and higher harmonic modes of the microwave resonator. We focus in this study on the first harmonic mode of the resonator as the sensitive microwave detection setup consisting of cryogenic circulators and low-noise amplifiers are optimized for frequencies around $5\,\giga\hertz$.

\section{Conventional ESR}

\begin{figure}[tb]
 \includegraphics[]{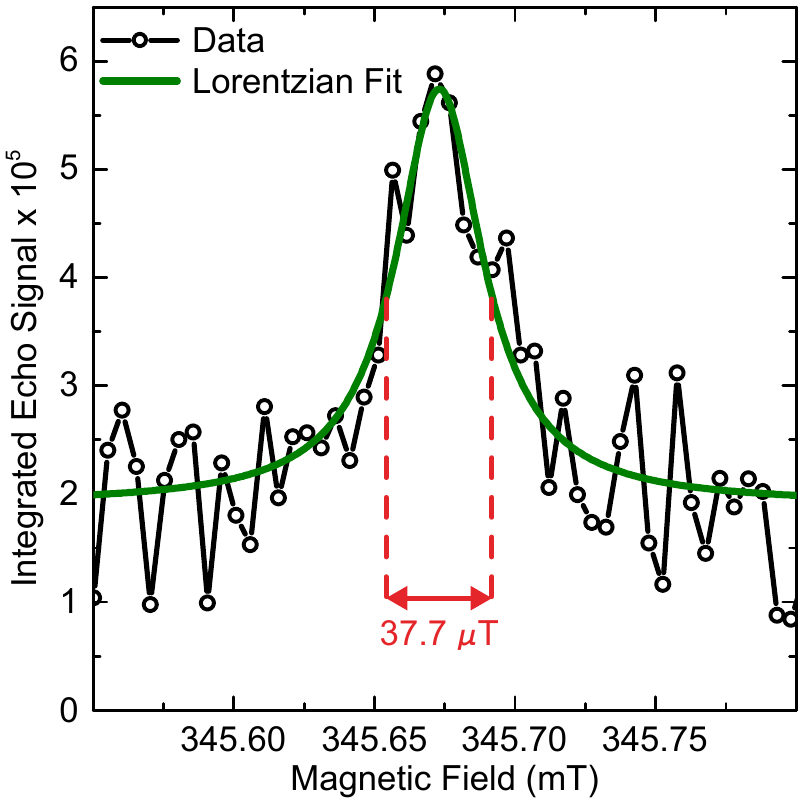}
 \caption{Echo detected magnetic field sweep of the high field spin resonance. \label{S2}}
 \end{figure}
For a comparison of the results from our millikelvin ESR setup we measured similarly prepared pieces of phosphorus-doped $^{28}$Si crystals in a commercial Bruker Elexsys E580 X-band spectrometer. Figure\,\ref{S2} shows the high field spin resonance acquired by a echo detected field sweep at a temperature of $8\,\kelvin$. Here, we applied at each magnetic field point a Hahn echo pulse sequence and detected the integrated echo signal via a magnitude detection method similar to the method discussed in Ref.\,\onlinecite{S_Tyryshkin2003}. By fitting a lorentzian function to the resonance peak we find a full width at half maximum of $\Delta B_{\text{Bruker}}=37.7\,\mu\tesla$. The magnetic linewidth translates to a spin loss rate $\gamma_{\text{Bruker}}/2\pi=\frac{g_{\text{eP}}\mu_{\text{B}}}{2h}\Delta B_{\text{Bruker}}=530\,\kilo\hertz$. For comparison, the loss rate extracted from the millikelvin measurements in the main text of $\gamma/2\pi=1.40\,\mega\hertz$ is nearly a factor of three greater. Therefore, the 
discrepancy has to origin from external inhomogeneities, which broaden the homogeneous spin resonance line.

\section{Population Probability}

In this section, we calculate the thermal polarization of the two electron spin resonance transitions present in phosphorus-doped silicon. The probabilty $p_i$ that a given energy state $E_i$ is occupied at a temperature $T$ is given by \cite{S_Blundell2001}
\begin{equation}
 p_i = \frac{e^{-E_i/k_{\mathrm{B}}T}}{Z},
\end{equation}
where $k_{\textmd{B}}$ is the Boltzman constant. The probabilty is normalized to the single particle partition function $Z$ which is the sum of all Boltzman factors of all avaliable energy states $E_i$
\begin{equation}
 Z = \sum_i e^{-E_i/k_{\mathrm{B}}T}.
\end{equation}
Assuming unstrained silicon the possible energy levels for phosphorus donors in silicon are given by the Hamiltonian
\begin{equation}
 H = \frac{g_{\mathrm{e}}\mu_{\mathrm{B}}}{\hbar}\textbf{B}\hat{\textbf{S}} + \frac{g_{\mathrm{N}}\mu_{\mathrm{N}}}{\hbar}\textbf{B}\hat{\textbf{I}} + \frac{A}{\hbar^2}\hat{\textbf{S}}\hat{\textbf{I}},
 \label{eqn:hamilton}
\end{equation}
where the first two terms are the electron and nuclear Zeeman interaction, respectively, and the last term represents the hyperfine interaction between the donor nuclear and electron spins. $g_{\mathrm{e}}$ $(g_{\mathrm{N}})$ represent the electron (nuclear) g-factor, $\mu_{\text{B}}$ $(\mu_{\text{N}})$ the Bohr (nuclear) magneton, $A$ the isotropic hyperfine constant and $\textbf{B}$ the magnetic field. We define the electron and nuclear spin operators $\hat{\textbf{S}}$ and $\hat{\textbf{I}}$ in their individual eigenbasis to
\begin{eqnarray}
 &\hat{\textbf{S}} = \frac{\hbar}{2}
 \begin{pmatrix}
  \hat{\sigma}_{\textmd{x}} \\
  \hat{\sigma}_{\textmd{y}} \\
  \hat{\sigma}_{\textmd{z}}
 \end{pmatrix}, \qquad 
 \hat{\textbf{I}} = \frac{\hbar}{2}
 \begin{pmatrix}
  \hat{\sigma}_{\textmd{x}} \\
  \hat{\sigma}_{\textmd{y}} \\
  \hat{\sigma}_{\textmd{z}}
 \end{pmatrix},
\end{eqnarray}
where $\hat{\sigma}_{\textmd{x}}$, $\hat{\sigma}_{\textmd{y}}$ and $\hat{\sigma}_{\textmd{z}}$ are the Pauli spin matrices for spin 1/2. With zero applied magnetic field the solution is a symmetric spin triplett state and an antisymmetric spin singlet state separated in energy by the hyperfine interaction. By applying a magnetic field the degeneracy is lifted by the Zeeman interaction and, in the limit of high fields, four distinct energy levels are present in the spin system.

\begin{figure}
 \includegraphics[]{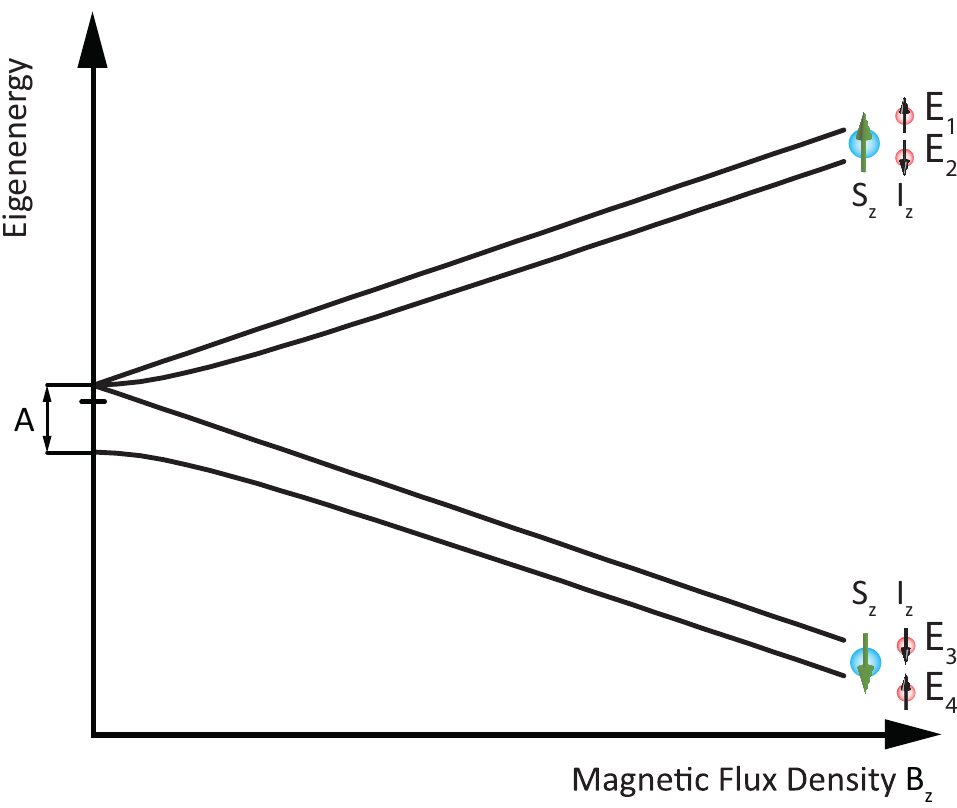}
 \caption{Schematic Breit-Rabi diagram for Si:P. \label{S3}}
 \end{figure}
We calculate the thermal population probabilties via the eigenenergies of \eqref{eqn:hamilton}. In matrix representation and assuming $\textbf{B}\,=\,(0,0,B_{\text{z}})$, \eqref{eqn:hamilton} takes the form
\begin{equation}
 \scriptstyle H =
 \begin{pmatrix}
  \scriptstyle \frac{A}{4} + \frac{B_{\mathrm{z}}}{2} \left( g_{\mathrm{e}}\mu_{\mathrm{B}} + g_{\mathrm{N}}\mu_{\mathrm{N}} \right) & \scriptstyle 0 & \scriptstyle 0 & \scriptstyle 0 \\
  \scriptstyle 0 & \scriptstyle -\frac{A}{4} + \frac{B_{\mathrm{z}}}{2} \left( g_{\mathrm{e}}\mu_{\mathrm{B}} - g_{\mathrm{N}}\mu_{\mathrm{N}} \right) & \scriptstyle \frac{A}{2} & \scriptstyle 0 \\
  \scriptstyle 0 & \scriptstyle \frac{A}{2} & \scriptstyle -\frac{A}{4} + \frac{B_{\mathrm{z}}}{2} \left( -g_{\mathrm{e}}\mu_{\mathrm{B}} + g_{\mathrm{N}}\mu_{\mathrm{N}} \right) & \scriptstyle 0 \\
  \scriptstyle 0 & \scriptstyle 0 & \scriptstyle 0 & \scriptstyle \frac{A}{4} - \frac{B_{\mathrm{z}}}{2} \left( g_{\mathrm{e}}\mu_{\mathrm{B}} + g_{\mathrm{N}}\mu_{\mathrm{N}} \right)
 \end{pmatrix}
 .
\end{equation}
The eigenenergies of the system are
\begin{eqnarray}
 E_1 &=& \frac{A}{4} + \frac{B_{\mathrm{z}}}{2} \left( g_{\mathrm{e}}\mu_{\mathrm{B}} + g_{\mathrm{N}}\mu_{\mathrm{N}} \right), \\
 \nonumber \\
 E_2 &=& -\frac{A}{4} + \frac{1}{2}\sqrt{ A^2 + B_{\mathrm{z}}^2 \left( g_{\mathrm{e}}\mu_{\mathrm{B}} - g_{\mathrm{N}}\mu_{\mathrm{N}} \right)^2 }, \\
 \nonumber \\
 E_3 &=& \frac{A}{4} - \frac{B_{\mathrm{z}}}{2} \left( g_{\mathrm{e}}\mu_{\mathrm{B}} + g_{\mathrm{N}}\mu_{\mathrm{N}} \right) \text{ and} \\
 \nonumber \\
 E_4 &=& -\frac{A}{4} - \frac{1}{2}\sqrt{ A^2 + B_{\mathrm{z}}^2 \left( g_{\mathrm{e}}\mu_{\mathrm{B}} - g_{\mathrm{N}}\mu_{\mathrm{N}} \right)^2 }.
\end{eqnarray}
For phosphorus in silicon $g_{\mathrm{e}}=1.9985$, $g_{\mathrm{N}}=2.2632$ and $A/h=117.53\,\mega\hertz$. \cite{S_Stone2005, S_Feher1959} The resulting Breit-Rabi diagram is shown in Figure \,\ref{S3}.

\begin{figure}[!ht]
 \includegraphics[]{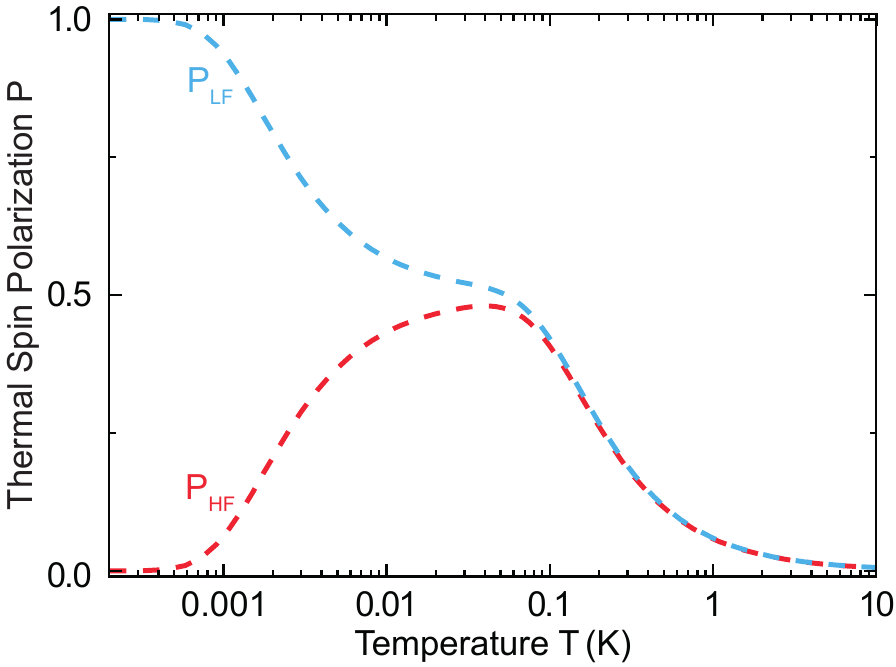}
 \caption{Spin polarization for the two allowed electron spin resonance transitions. \label{S4}}
\end{figure}

The corresponding population probabilities in thermal equilibrium are given by
\begin{eqnarray}
 p_1 = \frac{e^{-E_1/k_{\text{B}}T}}{\sum_{i=1..4} e^{-E_i/k_{\text{B}}T}}, \\
 \nonumber \\
 p_2 = \frac{e^{-E_2/k_{\text{B}}T}}{\sum_{i=1..4} e^{-E_i/k_{\text{B}}T}}, \\
 \nonumber \\
 p_3 = \frac{e^{-E_3/k_{\text{B}}T}}{\sum_{i=1..4} e^{-E_i/k_{\text{B}}T}} \text{ and} \\
 \nonumber \\
 p_4 = \frac{e^{-E_4/k_{\text{B}}T}}{\sum_{i=1..4} e^{-E_i/k_{\text{B}}T}}.
\end{eqnarray}
From these we deduce the electron spin polarization $P(T,B_{\textmd{z}})$ of the low field (LF) and the high field (HF) electron spin transitions ($E_1 \rightarrow E_4$ and $E_2 \rightarrow E_3$)
\begin{eqnarray}
 P_{\text{LF}}(T,B_{\text{z}}) = \left| p_1 - p_4 \right| \text{ and} \\
 \nonumber \\
 P_{\text{HF}}(T,B_{\text{z}}) = \left| p_2 - p_3 \right| \text{, respectively.}
\end{eqnarray}
The polarization of the two electron spin resonance transitions is displayed in Figure \,\ref{S4} for a fixed magnetic field of $176.5\,\milli\tesla$, resembling the conditions of our experiment.

\section{Estimate of the Effective Coupling Strength}
For a first estimation of the effective coupling strength $g_{\mathrm{eff}}$, we start with \cite{S_Huebl2013}
\begin{equation}
 g_{\mathrm{eff}} = \frac{g_{\mathrm{s}}\mu_{\mathrm{B}}}{2\hbar} B_1 \sqrt{N}.
\end{equation}
for spin 1/2 and a linear polarized microwave field \cite{S_Amsuess2012, S_Scully1997}. Here, $\hbar$ is the reduced Planck constant, $\mu_{\mathrm{B}}$ is the electron Bohr magneton, $g_{\mathrm{s}}$ is the electron g-factor of the spins and $N$ is the number of spins interacting with the oscillatory vacuum magnetic field $B_1$. We normalize the vacuum $B_1$ field to the ground state energy density of the resonator
\begin{equation}
 \frac{1}{2} \frac{\hbar\omega_{\mathrm{r}}}{2} = \frac{1}{2\mu_0} \int B_1^2 dV = \frac{1}{2\mu_0} B_1^2 V_{\mathrm{m}},
 \label{normB1}
\end{equation}
with the magnetic constant $\mu_0$ and $\omega_{\mathrm{r}}$ and $V_{\mathrm{m}}$ are the resonance frequency and the mode volume of the resonator, respectively. Note, that the energy stored in the resonator field is equally distributed in the magnetic and electric field. This is taken into account by the factor of $1/2$ on the left hand side of (\ref{normB1}) \cite{S_Schoelkopf2008}. The collective coupling strength $g_{\mathrm{eff}}$ yields
\begin{equation}
 g_{\mathrm{eff}} = \frac{g_{\mathrm{s}}\mu_{\mathrm{B}}}{2\hbar} B_1 \sqrt{\frac{N}{2}} = \frac{g_{\mathrm{s}}\mu_{\mathrm{B}}}{2\hbar}\sqrt{ \frac{\mu_0\rho\hbar\omega_{\mathrm{r}}V}{2V_{\mathrm{m}}} \frac{1}{2} } = \frac{g_{\mathrm{s}}\mu_{\mathrm{B}}}{2\hbar}\sqrt{ \frac{\mu_0\rho\hbar\omega_{\mathrm{r}}}{2} \frac{\eta}{2} },
\end{equation}
where $\rho$ is the phosphorus donor spin concentration and $V$ is the resonator mode volume filled with the doped silicon sample. The ratio $V/V_{\mathrm{m}}$ can be understood in terms of the filling factor $\eta$ of the resonator. We obtain $g_{\mathrm{eff}}/(2\pi) = 4.48\,\mega\hertz$ for the optimal condition that the complete upper half of the CPWR mode volume is filled, which corresponds to $\eta = 0.5$ and uses a spin density $\rho=1\times10^{17}\,\centi\meter^{-3}$.

Phosphorus donors in silicon show two electron spin resonance transitions due to the finite hyperfine interaction between the electron and nuclear spin of the phosphorus donor. At a temperature of $50\,\milli\kelvin$ the electron spins are fully polarized and the nuclear spin polarization is about $2.5\,\%$. For the specific case of phosphorus donors in silicon we can thus safely approximate that each of the two hyperfine split ESR transitions contains $50\,\%$ of the total spin density $\rho$. Therefore, the number of transitions (2) reduces $g_{\mathrm{eff}}$  by an additional factor of $\sqrt{2}$ yielding $g_{\mathrm{eff}}/(2\pi)=3.17\,\mega\hertz$ for the individual transition. A value which is larger than the experimentally observed values of $1.07\,\mega\hertz$ and $1.13\,\mega\hertz$.

For the more realistic situation of a finite gap of width $d$ between the $^{28}$Si crystal and the coplanar microwave resonator we calculate $g_{\mathrm{eff}}$ numerically. We start by computing the microwave $B_1$ field distribution according to Ref.\,\onlinecite{S_Simons1982}, where we again normalize the $B_1$ field amplitude to the ground state energy density of the resonator given in (\ref{normB1}). Hereby, we obtain a position-dependent coupling strength $g_{0}(\textbf{r}_i)$. In total we obtain for the effective coupling \cite{S_Sandner2012}
\begin{equation}
 g_\mathrm{eff} = \sqrt{\frac{1}{2} \sum_{i=1}^{\mathrm{N}} |g_{0}(\textbf{r}_i)|^2} = \frac{g_{\mathrm{s}}\mu_{\mathrm{B}}}{2\hbar} \sqrt{\frac{1}{2} \sum_{i=1}^{\mathrm{N}} |B_1(\textbf{r}_i)|^2}.
\end{equation}
The index $i$ refers to each individual spin contained inside the phosphorus-doped $^{28}$Si crystal. The calculation is performed on a discrete lattice with lattice constant $a$, where at each lattice site a spin is located given by the vector $\textbf{r}_i$. Assuming a homogeneous spin distribution in the sample, we determine the lattice constant by the mean inter-particle distance to $26.73\,\nano\meter$ for a phosphorus donor concentration of $1\,\times\,10^{17}\,\centi\meter^{-3}$. The numerical $B_1$ field calculation is performed for the cross-section of the CPW \cite{S_Simons1982}, which we place in the $yz$-plane with the $z$-coordinate being normal to the CPW surface. For the $x$-coordinate along the length $l$ of the CPWR, we modulate the $yz$-field with a sinusoidal reflecting the shape of the first harmonic mode of the resonator,
\begin{equation}
 g_\mathrm{eff} = \frac{g_{\mathrm{s}}\mu_{\mathrm{B}}}{2\hbar} \sqrt{\frac{1}{2} \sum_{i=1}^{\mathrm{N}} |B_1(\textbf{r}_i)|^2} = \frac{g_{\mathrm{s}}\mu_{\mathrm{B}}}{2\hbar} \sqrt{\frac{1}{2} \sum_{i=1}^{\mathrm{N}} \left[ (B_{\mathrm{y,}i}^2 + B_{\mathrm{z,}i}^2) \, \left|\sin\left(\frac{2\pi l}{x_i}\right)\right|^2 S \right] }.
\end{equation}
For electron spin resonance the oscillatory magnetic field $B_1$ needs to be perpendicular to the static magnetic field $B_0$. This condition is not fulfilled on the whole length $l$ of the resonator (cf. Fig.\,\ref{S1}). Thus, for the calculation we only consider the parts of the resonator where $B_1$ is perpendicular to $B_0$ corresponding to locations where the center line of the CPWR is aligned in parallel with the static magnetic field $B_0$ (cf. Fig.\,1\,(a) of the main text), which is accounted for by the piecewise defined function
\begin{equation}
 S =
  \begin{cases}
    1  & \quad \text{if CPW} \enspace || \enspace B_0\\
    0  & \quad \text{else}\\
  \end{cases} .
\end{equation}
For a gap of width of $d = 12.5\,\micro\meter$ we obtain $g_{\mathrm{eff}}/(2\pi)= 1.10\,\mega\hertz$, which is in good agreement with the experimentally determined values.


\begin{thebibliography}{40}
\bibitem [{cou()}]{coupling_comment}%
  \BibitemOpen
  \href@noop {} {}\bibinfo {note} {This is taken into account by the 4 in the
  denominator of $g_{\mathrm{eff,th}}$}\BibitemShut {NoStop}%
\end{thebibliography}
\end{document}